# Particle-hole asymmetric superconducting coherence peaks in overdoped cuprates

Changwei Zou[1,*], Zhenqi Hao[1,*], Xiangyu Luo[2], Shusen Ye[1], Qiang Gao[2], Xintong Li[1], Miao Xu[1], Peng Cai[3], Chengtian Lin[4], Xingjiang Zhou[2], Dung-Hai Lee[5,6], Yayu Wang[1,7†]

[1]*State Key Laboratory of Low Dimensional Quantum Physics, Department of Physics, Tsinghua University, Beijing 100084, P. R. China*

[2]*Beijing National Laboratory for Condensed Matter Physics, Institute of Physics, Chinese Academy of Sciences, Beijing 100190, P. R. China*

[3]*Department of Physics, Renmin University, Beijing 100872, P. R. China*

[4]*Max Planck Inst Solid State Res, Heisenbergstr 1, D-70569 Stuttgart, Germany*

[5]*Department of Physics, University of California at Berkeley, Berkeley, CA 94720.*

[6]*Materials Sciences Division, Lawrence Berkeley National Laboratory, Berkeley, CA 94720.*

[7]*Frontier Science Center for Quantum Information, Beijing 100084, P. R. China*

[*]These authors contributed equally to this work.

[†]yayuwang@tsinghua.edu.cn

**Abstract:**

To elucidate the superconductor to metal transition at the end of superconducting dome, the overdoped regime has stepped onto the center stage of cuprate research recently. Here, we use scanning tunneling microscopy to investigate the atomic-scale electronic structure of overdoped trilayer and bilayer $Bi_2Sr_2Ca_{n-1}Cu_nO_{2n+4+\delta}$ cuprates. At low energies the spectroscopic maps are well described by dispersive quasiparticle interference patterns. However, as the bias increases to the superconducting coherence peak energy, a virtually non-dispersive pattern with $\sqrt{2}a \times \sqrt{2}a$ periodicity emerges. Remarkably, the position of the coherence peaks exhibits evident particle-hole asymmetry which also modulates with the same period. We propose that this is an extreme quasiparticle interference phenomenon, caused by pairing-breaking scattering between flat anti-nodal Bogoliubov bands, which is ultimately responsible for the superconductor to metal transition.

In the past three and half decades, research on cuprate high-temperature superconductors primarily focuses on the underdoped and optimally doped regions of the phase diagram (*1*). Striking phenomena such as pseudogap and strange metal "normal" states together with the existence of a plethora of nearby electronic orders cause the pairing mechanism question difficult to answer. It is until relative recently this research has branched into studying the overdoped regime. Specific heat (*2*), penetration depth (*3*), and optical conductivity (*4*) measurements done on $La_{2-x}Sr_xCuO_4$ films have shown that there is a large portion of normal state Drude weight which remains uncondensed in as $T \to 0$. As a consequence the superfluid density is low, and eventually vanishes at the quantum critical doping of the superconductor to metal transition. On a different front, recent angle resolved photoemission spectroscopy (ARPES) study on overdoped $Bi_2Sr_2CaCu_2O_{8+\delta}$ (Bi-2212) shows clear evidence that Cooper pairs have already formed at temperature significantly above $T_c$ (*5*). These phenomena are in sharp contrast to both transport (*6, 7*) and spectroscopic (*8, 9*) studies suggesting the normal state become "less" strange with doping. This dichotomy between the behaviors of the superconducting (SC) versus the normal state sets the stage for the recent research on the overdoped side of the cuprate phase diagram.

As far as scanning tunneling microscopy (STM) is concerned, lots of works have been done in the optimal and underdoped regime in the literature (*10*). When the bias voltage is in the range of the nodal "Fermi arc", Bogoliubov quasiparticle (BQP) interference patterns predicted by the octet model describes the data well (*11, 12*). When the bias is increased to the expected coherence peak (i.e. anti-nodal BQP excitation) energy, a disordered translation (*13, 14*) and rotation symmetry breaking tunneling patterns (*15, 16*) emerges. Such patterns are somewhat correlated with that found in the core of SC vortices (*17*).

Relatively less work was done on the overdoped side. First, the pseudogap vanishes, and both ARPES and STM works show an untruncated Fermi surface (*8, 18*). In addition, ARPES find well-defined anti-nodal "quasiparticle" peaks (*9, 19*). In real space the crystal symmetry breaking patterns are no longer visible at the expected energy of the coherence peak (*18*). These facts motivate one to think the BQP excitations at the antinode should continue to reveal the quasiparticle interference (QPI) phenomenon. However, a red flag against this line of thought is raised by the fact that ARPES found the BQP dispersion parallel to the Brillouin zone boundary is extremely flat (*20, 21*). In fact, that dispersion width is comparable to, if not smaller than, the value of the SC gap. Therefore, we have a situation in apparent contradiction to the BCS limit. Moreover, one would expect the flat dispersion to give rise to large density of state (DOS) hence enhance the antinode-antinode disorder scattering. Such scattering takes place between momentum regions where the *d*-wave SC gap function assumes opposite sign, hence it should be pair breaking. In view of the above consideration, it is much less clear what the STM pattern should be at the coherence peak (i.e. SC gap edge) energy.

The trilayer $Bi_2Sr_2Ca_2Cu_3O_{10+\delta}$ (Bi-2223) represents an ideal system to study the local electronic structure of overdoped cuprates. Partly it is because Bi-2223 has the highest $T_c$ in the Bi-based family (*22*), making the detection of low energy excitations more accurate. In addition, Bi-2223 crystals can be easily cleaved to expose a clean BiO surface (Fig. 1A inset), thus is highly suitable for STM studies (*23, 24*). As Bi-2212, upon overdoping the pseudogap vanishes in Bi-2223 (*25, 26*). In the following we discuss the atomic scale electronic structure of Bi-2223 with varied dopings.

Figure 1A displays the topographic image of a slightly underdoped Bi-2223 with $T_c = 110$ K (Sample-1 of the six Bi-2223 samples studied in this work, see Materials and Methods). Here the Bi atoms, separated by lattice constant $a \sim 0.38$ nm, are clearly resolved. In Fig. 1B we plot the spatially averaged d$I$/d$V$ spectra sorted by the local gap size $\Delta$. It shows that spectra with smaller $\Delta$ have sharper SC coherence peaks while those with large $\Delta$ are influenced by the pseudogap, and exhibit broad peaks. In order to eliminate the influence of set-point and static charge order, a common practice is to plot the ratio of conductance maps at $\pm E$, namely, $Z(\mathbf{r}, E) = g(\mathbf{r}, E)/g(\mathbf{r}, -E)$ (*27*), as displayed in supplementary Fig. S1. The low-energy states exhibit well-defined QPI patterns in the bias range covered by the green rectangle in Fig.1B. An example of which is given by the "Z map" at $E = 20$ meV in Fig. 1C. To achieve better momentum-space resolution, the Fourier-transformed (FT) map $Z(\mathbf{q}, 20$ meV$)$ in Fig. 1D is acquired in a larger field of view (see Fig. S2). It reveals the seven wavevectors expected by the octet-model as illustrated in Fig. 1G. To further reveal the phase of the SC gap function, we apply the phase referenced (PR) technique that has been developed on cuprates and iron pnictides (*28-30*). Shown in Fig. 1H is the PR map defined by $PR(\mathbf{q},E)=|g(\mathbf{q},-E)|\cos(\delta\varphi)$, where $|g(\mathbf{q},-E)|$ is the amplitude of $g(\mathbf{q},-E)$ and $\delta\varphi=\varphi_{\mathbf{q},-E}-\varphi_{\mathbf{q},+E}$ is the phase difference between $g(\mathbf{q},-E)$ and $g(\mathbf{q},+E)$. The negative PR signals (blue) at $E = 20$ meV appear at $\mathbf{q}_2$, $\mathbf{q}_3$, $\mathbf{q}_6$, and $\mathbf{q}_7$, which perfectly match the sign-reversing scattering between the eight "hot spots" in Fig. 1G for nodal BQP (*30*).

When $E$ is increased towards the gap edge, the QPI patterns are replaced by a "glassy stripe" patterns (Fig. 1E) highly analogous to that found in underdoped Bi-2212 and $Ca_{2-x}Na_xCuO_2Cl_2$ (*31*). To minimize the influences of structural and chemical disorders, the energies in Fig. 1E are scaled to the local gap sizes with $\varepsilon = E(\mathbf{r})/\Delta(\mathbf{r}) = 1$, following the practice in Ref. (*32*). In the FT image in Fig. 1F, $\mathbf{q}_1^*$ and $\mathbf{q}_5^*$ label the non-dispersive wavevectors corresponding to the "stripe" order (*31, 32*), which has been attributed to the incoherent anti-nodal states intimately linked to the pseudogap.

Next, we examine the electronic structure of a heavily overdoped Bi-2223 (Sample-5) with $T_c = 109$ K. The topography shown in Fig. 2A is similar to that in Sample-1, but the gap sizes shown in Figs. 2B and 2C are considerably smaller, with an average $\Delta \sim 30$ meV. There still exists an anti-correlation between local $\Delta$ and the coherence peak height. However, all the spectra display sharp coherence peaks consistent with the absence of the pseudogap. The $Z$ map taken at low energy $E = 10$ mV exhibits clear QPI patterns (Figs. 2D and 2G) obeying the prediction of the octet model. The maximum energy in which QPI is observed is reduced to ~ 20 meV, shown as the green area in Fig. 2C and supplementary Fig. S3.

At bias close to the gap edge, the dispersive (i.e. bias-dependent) QPI patterns is replaced by a nearly non-dispersive (see Figs. S3 and S4 for the complete bias evolution), short-range, $\sqrt{2}a \times \sqrt{2}a$ pattern oriented along the diagonal direction of $CuO_2$ plaquette, as shown by the $Z$ map at $\varepsilon = 0.75$ in Fig. 2E. The FT map in Fig. 2H reveals four prominent, albeit broad, wavevectors around ($\pm\pi/a$, $\pm\pi/a$), which are denoted as $\mathbf{q}_a$ and $\mathbf{q}_b$. This phenomenon is not hard to understand – it corresponds to the OPI pattern associated with the disorder scattering between antinodes (the black arrow in Fig. 2F). The fact that such scattering is pair-breaking is confirmed by the PR analysis shown in Fig. 2H, in which all the four wavevectors are associated with scattering between regions with opposite sign of the gap function. This pair breaking scattering is further supported by the anti-correlation between the strength of the nanoscale $\sqrt{2}a \times \sqrt{2}a$ patches (Fig. 2E) and the heterogeneous local 2Δ (Fig. 2B), namely, in region with smaller 2Δ has stronger $\sqrt{2}a \times \sqrt{2}a$ modulation.

A closer examination of the spatially resolved d$I$/d$V$ spectra unveils another highly surprising phenomenon. Figure 3A is a zoomed-in topography superposed with the Cu lattice, in which the Cu atoms are divided into two sublattices marked with red and blue dots. Figure 3B displays the $Z$ map at ε = 0.75, which reveals that the $\sqrt{2}a \times \sqrt{2}a$ pattern is the period of the blue/red Cu sites. In Fig. 3C we plot the d$I$/d$V$ spectra along the two cuts highlighted in yellow in Figs. 3A and 3B. The bottom of each d$I$/d$V$ spectrum is pinned at zero bias, and the low energy lineshape is ±$E$ symmetric and homogeneous. However, the positions of the SC coherence peaks oscillate between $(-\Delta_L, +\Delta_S)$ and $(-\Delta_S, +\Delta_L)$ from the blue to the red sublattice. Here $\Delta_L \neq \Delta_S$ hence the coherence peaks are not ±$E$ symmetric. The periodic "swing" of the position of the coherence peaks can be visualized in the "gap asymmetry map" defined by $\frac{\delta\Delta}{\Delta}(\mathbf{r}) = 2\frac{\Delta_L(\mathbf{r}) - \Delta_S(\mathbf{r})}{\Delta_L(\mathbf{r}) + \Delta_S(\mathbf{r})}$. As shown in Fig. 3D, the δΔ/Δ map has the same $\sqrt{2}a \times \sqrt{2}a$ periodicity as the $Z$ map in Fig. 2E (see also Fig. S7).

To test the universality of the above phenomenon, we repeat the measurements on double-layer Bi-2212 including an underdoped sample with $T_c$ = 82 K, (denoted as UD82K) and two overdoped samples denoted as OD81K and OD71K. The results are nearly identical but the $\sqrt{2}a \times \sqrt{2}a$ modulations are weaker than that in Bi-2223, as displayed by the $Z$ map at ε = 0.75 of the OD71K sample in Fig. 4B (see the complete data in supplementary Fig. S6). We note that the wavevectors corresponding to $\sqrt{2}a \times \sqrt{2}a$ are also present in previous STM work on overdoped Bi-2212 (*18*), but the real-space pattern was not identified and the spectral features were not analyzed (see Fig. S8).

To quantify the correlation between the asymmetry parameter δΔ/Δ and 2Δ, we use a standard short-time FT method (see supplementary Fig. S9). Figure 4A summarizes the spatially averaged δΔ/Δ of five overdoped Bi-2223 (red triangles) and two overdoped Bi-2212 (blue triangles) samples as a function of averaged 2Δ (the latter is anticorrelated with $p$, see supplementary Section C). When the pseudogap vanishes with overdoping, the $\sqrt{2}a \times \sqrt{2}a$ modulation of the SC coherence peak with particle-hole asymmetry emerges and becomes stronger. The black dots show the anti-correlation between the local δΔ/Δ and the local 2Δ, they are adopted from Bi-2223 Sample-5. Under the same 2Δ value, the δΔ/Δ amplitudes of the Bi-2212 samples are weaker than that in Bi-2223, which is probably why this phenomenon was missed in previous STM studies of Bi-2212. This trend also explains why such modulations are not found in the SC phase of overdoped $Bi_2Sr_2CuO_{6+\delta}$ (Bi-2201), in which the superconductivity is much weaker and the pseudogap persists to higher dopings (*33*). We note that in the heavily overdoped non-SC Bi-2201, a $\sqrt{2}a \times \sqrt{2}a$ pattern was observed (*34*). It can be easily described by the disorder induced scattering of normal "quasiparticles" (*34*), and is not related to superconductivity.

The main puzzle left to explain is why isn't the coherence peak ±$E$ symmetric? The answer lies in strong antinode-antinode scattering triggered by disorder. Such scattering is enhanced by the large joint DOS associated with the flat dispersion along the Brillouin zone face, as is evidenced by the high SC coherence peaks in the d$I$/d$V$ curves of overdoped cuprates (Figs. 2C, S5 and S6). In Fig. 4C we show a caricature conveying our interpretation of the origin of the asymmetry. In the perfectly flat band limit, let the BQP eigen energy level at the antinode be represented by the black line segments in Fig. 4C. The inter-antinode scattering mixes the energy levels associated with the $\mathbf{k}_1 = (\pi, 0)$ and $\mathbf{k}_2 = (0, \pi)$ with the BQP eigen wavefunctions $\varphi_{\mathbf{k}_1}(\mathbf{r}) = \frac{1}{\sqrt{2}}\begin{pmatrix}1\\1\end{pmatrix}e^{i\mathbf{k}_1 \cdot \mathbf{r}}$ and $\varphi_{\mathbf{k}_2}(\mathbf{r}) = \frac{1}{\sqrt{2}}\begin{pmatrix}1\\-1\end{pmatrix}e^{i\mathbf{k}_2 \cdot \mathbf{r}}$ for, say, the $E < 0$ levels. Here the upper/lower component of the wavefunction denotes the particle/hole component of the BQP. Note the relative minus sign between the hole component is due to the opposite sign of the SC gap function at $\boldsymbol{k}_1$ and $\boldsymbol{k}_2$. Similarly, the eigen wavefunctions associated with the $E > 0$ energy level are $\chi_{\mathbf{k}_1}(\mathbf{r}) = \frac{1}{\sqrt{2}}\begin{pmatrix}1\\-1\end{pmatrix}e^{i\mathbf{k}_1 \cdot \mathbf{r}}$ and $\chi_{\mathbf{k}_2}(\mathbf{r}) = \frac{1}{\sqrt{2}}\begin{pmatrix}1\\1\end{pmatrix}e^{i\mathbf{k}_2 \cdot \mathbf{r}}$. Assuming the inter-antinode scattering potential is originated from scalar impurities, it has the form $V(\mathbf{r})\begin{pmatrix} 1 & 0 \\ 0 & -1 \end{pmatrix}$. This scattering potential only mixes $\varphi_{\mathbf{k}_1}(\mathbf{r})$ with $\varphi_{\mathbf{k}_2}(\mathbf{r})$ and $\chi_{\mathbf{k}_1}(\mathbf{r})$ with $\chi_{\mathbf{k}_2}(\mathbf{r})$. After the mixing, the energy level split as shown in Fig. 4C as the blue and red line segments. The minimum gap, $2\Delta_S$ is reduced by the scattering implying it is pair breaking. Depending on the phase, $e^{i\theta}$, of the matrix element $\int d^2\mathbf{r} e^{-i\mathbf{k}_2 \cdot \mathbf{r}} V(\mathbf{r}) e^{i\mathbf{k}_1 \cdot \mathbf{r}}$ we obtain the following spatial-modulating local DOS at the gap edges:

$$\text{LDOS}(E,\mathbf{r}) = [L(E+\Delta_L) + L(E+\Delta_S)] + \cos(\mathbf{q}\cdot\mathbf{r}+\theta)[L(E+\Delta_L) - L(E+\Delta_S)]$$

for $E < 0$, (1A)

$$\text{LDOS}(E,\mathbf{r}) = [L(E-\Delta_L) + L(E-\Delta_S)] - \cos(\mathbf{q}\cdot\mathbf{r}+\theta)[L(E-\Delta_L) - L(E-\Delta_S)]$$

for $E > 0$. (1B)

Here $L(E)$ is a Lorentzian centered at $E = 0$. Using the above formula we simulate a line cutting through the Cu sites (as that shown in Fig. 3B), which agrees well with the experimental observation qualitatively and suggests that observed phenomenon originates from an "extreme" version of the QPI. The adjective "extreme" refers to the fact that the strong inter-antinode scattering not only causes a superposition of the BQP wavefunctions associated with different momenta, it also has modified the BQP eigen energies (see Fig. 4C). Importantly, within this explanation the modified BQP energy levels remain $\pm E$ symmetric. The apparent asymmetry is due to the interference of the split BQP eigenfunctions.

Within our interpretation δΔ/Δ measures the strength of the pair-breaking scattering, and the patches size shown in Fig. 3D manifests the correlation length of the disorder scattering potential. Thus, they yield useful spectroscopic information about the disorder potential. The present result and interpretation bear strong relevance to the superconductor to metal transition discussed in the introduction. Recently theoretical study by Li *et al.* (*35*) has shown that even when the pairing interaction is held fixed, disorder induced antinode-antinode scattering can cause the system to spontaneously develop gap heterogeneity. In the limit of strong disorder this causes the superconductors to evolve into superconducting islands embedded in normal metals. Such heterogeneity has the effect of reducing the zero-temperature superfluid density, and eventually leads to the transition into a metallic state.

**Acknowledgements:** We thank T. K. Lee, T. Li, Q.H. Wang, Z.Y. Weng, and T. Xiang for helpful discussions. This work was supported by the NSFC grant No. 11534007, MOST of China grant No. 2017YFA0302900 and No. 2016YFA0300300, the Basic Science Center Project of NSFC under grant No. 51788104, NSFC Grant No. 11888101, and the Strategic Priority Research Program (B) of the Chinese Academy of Sciences (XDB25000000). DHL was supported by the U.S. Department of Energy, Office of Science, Basic Energy Sciences, Materials Sciences and Engineering Division, contract No. DE-AC02-05-CH11231 within the Quantum Materials Program (KC2202). DHL also acknowledge support from the Gordon and Betty Moore Foundation's EPIC initiative, Grant No. GBMF4545. This work is supported in part by the Beijing Advanced Innovation Center for Future Chip (ICFC).

**Figure Captions:**

**Figure 1. Nodal QPI and anti-nodal symmetry-breaking state in underdoped Bi-2223 (Sample-1).** (**A**) A topographic image taken with tunneling current $I$ = 20 pA and bias voltage $V_b$ = -250 mV. The inset shows the schematic crystal structure of Bi-2223. (**B**) Spatially averaged d$I$/d$V$ spectra sorted by different local SC gap size. The green shaded area marks the energy range in which the QPI phenomenon is apparent. (**C**) The $Z$ map in the same field-of-view (FOV) as (A) at $E$ = 20 meV, displaying well-defined QPI patterns. (**D**) The FT image of a large-area $Z$ map (Fig. S2) in the same sample at $E$ = 20 meV, which reveals characteristic QPI wavevectors. (**E**) The $Z$ map at $\varepsilon = E(\mathbf{r})/\Delta(\mathbf{r}) = 1$ in the same FOV as (A) with clear local breaking of translation and rotation symmetries. (**F**) The FT image of a large-area $Z$ map at $\varepsilon$ = 1 reveals non-dispersive wavevectors. (**G**) Schematic scattering process of the BQP on the Fermi arc with the seven wavevectors predicted by the octet model. The red shaded areas represent the anti-nodal region with pseudogap, and blue dots represent the "hot spots" with maximal joint DOS. (**H**) The phase-referenced analysis of the QPI pattern at $E$ = 20 meV, showing the sign-preserving (red) and sign-reversing (blue) scattering wavevectors. Note that the Bragg peaks (black circles) in (D), (F) and (H) are rotated by 45° compared to (C) and (E).

**Figure 2. The $\sqrt{2}a \times \sqrt{2}a$ modulation in an overdoped Bi-2223 (Sample-5).** (**A**) A topographic image taken with $I$ = 10 pA and $V$ = -150 mV. (**B**) The gap map in the same FOV as (A), where the local $2\Delta$ is the energy difference between the SC coherence peaks. (**C**) Spatially averaged d$I$/d$V$ spectra sorted by different local $\Delta$. The green shaded area marks the QPI energy range. (**D**) The $Z$ map at $E$ = 10 meV in the same FOV as (A), associated with the dispersive QPI patterns. (**E**) The $Z$ map at $\varepsilon$ = 0.75, displaying the short-range $\sqrt{2}a \times \sqrt{2}a$ modulation. (**F**) Schematic scattering process of the BQP between anti-nodal regions (blue shaded areas) connected by $\mathbf{q}$ = ($\pi$, $\pi$) as indicated by the arrow. (**G**) The FT of $Z$ map at $E$ = 10 meV in a larger FOV (Fig. S3). (**H**) The FT of the $Z$ map at $\varepsilon$ = 0.75, where $\mathbf{q}_a$ and $\mathbf{q}_b$ are the centers of the peak wavevectors. (**I**) The phase-referenced analysis of the $\sqrt{2}a \times \sqrt{2}a$ modulation at $\varepsilon$ = 0.75, showing that $\mathbf{q}_a$ and $\mathbf{q}_b$ are BQP scatterings between regions with opposite sign in the gap function, as expected in (F).

**Figure 3. Particle-hole asymmetric SC coherence peaks in Sample-5.** (**A**) The zoomed-in topography of the yellow square in Fig. 2A. The inset is the schematic structure of the $CuO_2$ plane, red and blue dots mark the two sublattices of the Cu sites beneath of the BiO surface. (**B**) The $Z(\mathbf{r}, \varepsilon = 0.75)$ map in the same FOV as (A). (**C**) The d$I$/d$V$ spectra along the two yellow cuts marked in (A) and (B), where the red and blue circles indicate the positions of the SC coherence peaks at Cu sites with the corresponding colored dots. (**D**) The gap asymmetry map in the same FOV as Fig. 2A.

**Figure 4. Summarized data and theoretical model of flat band scattering.** (**A**) Red triangles: the average $\delta\Delta/\Delta$ amplitude versus the average $2\Delta$ in the five overdoped Bi-2223 samples. Black circles: local $\delta\Delta/\Delta$ amplitude versus local $2\Delta$ values in Sample-5. Blue triangles: the average $\delta\Delta/\Delta$ amplitude in the two overdoped Bi-2212 samples. (**B**) The $Z(\mathbf{r}, \varepsilon = 0.75)$ map taken from the OD71K Bi-2212 sample. (**C**) Schematic of the inter-anti-nodal BQP scattering

process with **q** = (π, π), where the scattering in the same branch of BQP band is enhanced by the flat band under the d-wave gap symmetry, while the scattering between opposite branch vanishes. (**D**) Simulated real-space spectral evolution using Eq. (1A) and (1B), which shows the particle-hole asymmetry with $\sqrt{2}a$ periodicity. The simulating parameters in this model (see text) are: $\Gamma/\Delta = 0.14$, $\Delta_L/\Delta = 1.2$ and $\Delta_S/\Delta = 0.8$, here $\Gamma$ is the width of the Lorentzian and $\Delta$ is the average gap.

**References:**

1. B. Keimer, S. A. Kivelson, M. R. Norman, S. Uchida, J. Zaanen, *Nature* **518**, 179-186 (2015).
2. H.-H. Wen *et al.*, *Phys. Rev. B* **70**, 214505 (2004).
3. I. Bozovic, X. He, J. Wu, A. T. Bollinger, *Nature* **536**, 309-311 (2016).
4. F. Mahmood, X. He, I. Božović, N. P. Armitage, *Phys. Rev. Lett.* **122**, 027003 (2019).
5. Y. He *et al.*, https://arxiv.org/abs/2009.10932 (2020).
6. R. A. Cooper *et al.*, *Science* **323**, 603-607 (2009).
7. B. Vignolle *et al.*, *Nature* **455**, 952-955 (2008).
8. M. Plate *et al.*, *Phys. Rev. Lett.* **95**, 077001 (2005).
9. M. Hashimoto, I. M. Vishik, R.-H. He, T. P. Devereaux, Z.-X. Shen, *Nat. Phys.* **10**, 483-495 (2014).
10. Ø. Fischer, M. Kugler, I. Maggio-Aprile, C. Berthod, C. Renner, *Rev. Mod. Phys.* **79**, 353-419 (2007).
11. J. E. Hoffman *et al.*, *Science* **297**, 1148-1151 (2002).
12. Q. H. Wang, D. H. Lee, *Phys. Rev. B* **67**, 020511 (2003).
13. K. M. Lang *et al.*, *Nature* **415**, 412-416 (2002).
14. A. N. Pasupathy *et al.*, *Science* **320**, 196-201 (2008).
15. T. Hanaguri *et al.*, *Nature* **430**, 1001-1005 (2004).
16. C. V. Parker *et al.*, *Nature* **468**, 677-680 (2010).
17. J. E. Hoffman *et al.*, *Science* **295**, 466-469 (2002).
18. K. Fujita *et al.*, *Science* **344**, 612-616 (2014).
19. I. K. Drozdov *et al.*, *Nat. Commun.* **9**, 5210 (2018).
20. H. Li *et al.*, *Nat. Commun.* **9**, 26 (2018).
21. T. Valla, I. K. Drozdov, G. D. Gu, *Nat. commun.* **11**, 569 (2020).
22. H. Matsui *et al.*, *Phys. Rev. Lett.* **90**, 217002 (2003).
23. N. Jenkins *et al.*, *Phys. Rev. Lett.* **103**, 227001 (2009).
24. C. Zou *et al.*, *Phys. Rev. Lett.* **124**, 047003 (2020).

25. Y. Yamada *et al.*, *Phys. Rev. B* **68**, 054533 (2003).
26. A. Piriou, Y. Fasano, E. Giannini, O. Fischer, *Phys. Rev. B* **77**, 184508 (2008).
27. T. Hanaguri *et al.*, *Nat. Phys.* **3**, 865-871 (2007).
28. T. Hanaguri *et al.*, *Science* **323**, 923-926 (2009).
29. P. O. Sprau *et al.*, *Science* **357**, 75-80 (2017).
30. Q. Gu *et al.*, *Nat. Commun.* **10**, 1603 (2019).
31. Y. Kohsaka *et al.*, *Science* **315**, 1380-1385 (2007).
32. Y. Kohsaka *et al.*, *Nature* **454**, 1072-1078 (2008).
33. Y. He *et al.*, *Science* **344**, 608-611 (2014).
34. X. Li *et al.*, *New J. Phys.* **20**, 063041 (2018).
35. Z. Li, S. Kivelson, D.-H. Lee, https://arxiv.org/abs/2010.06091 (2020).

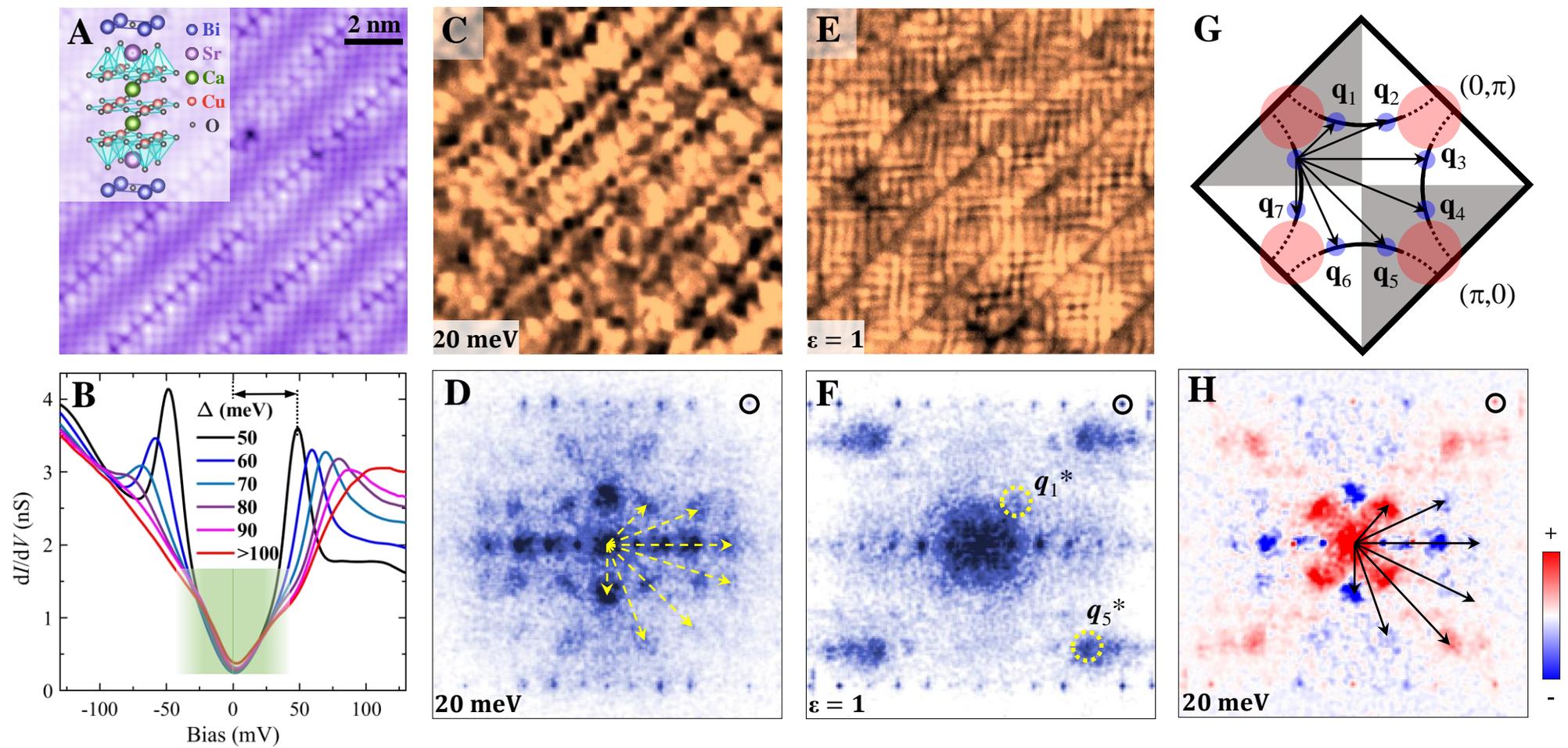

**Figure 1**

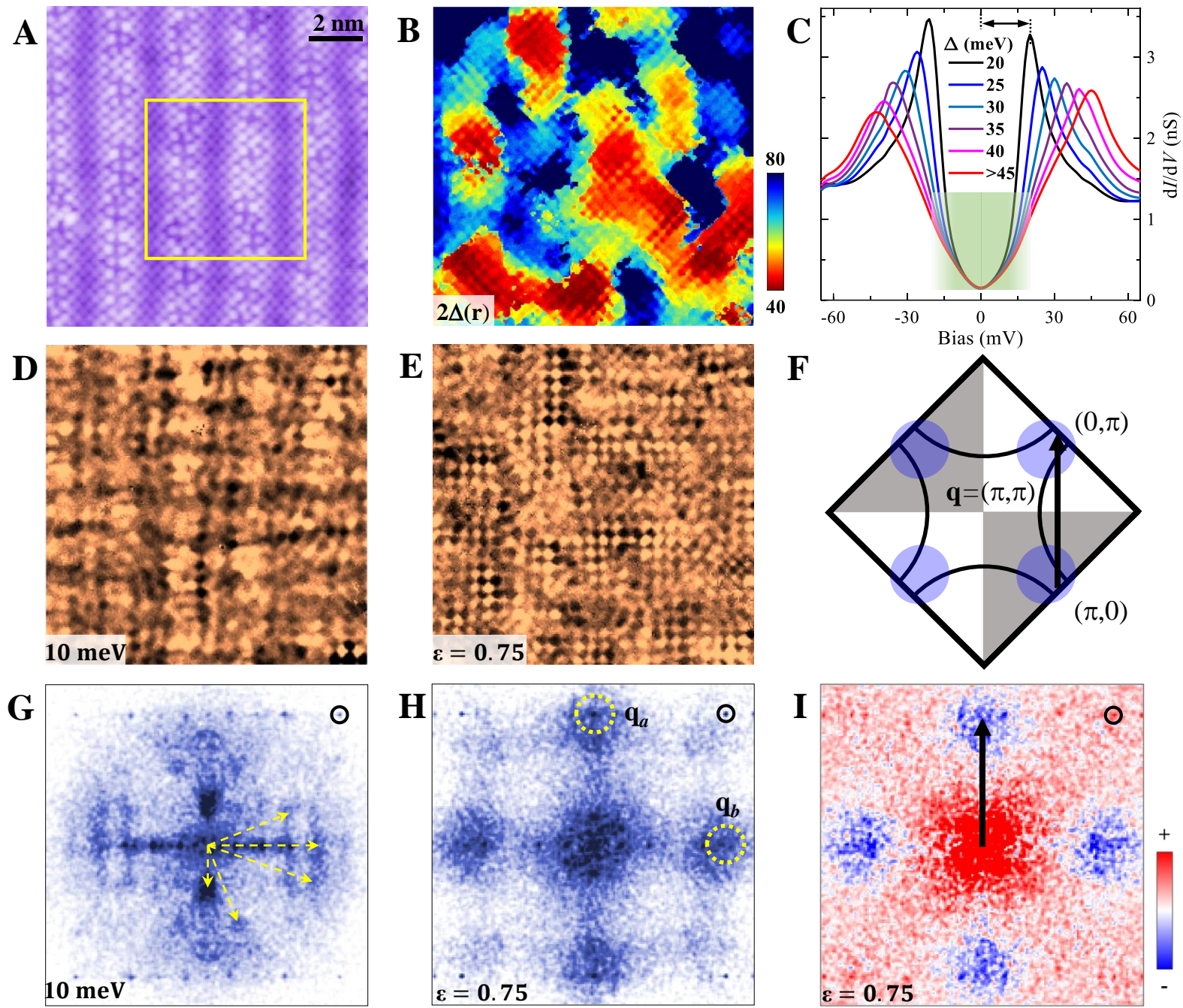

Figure 2

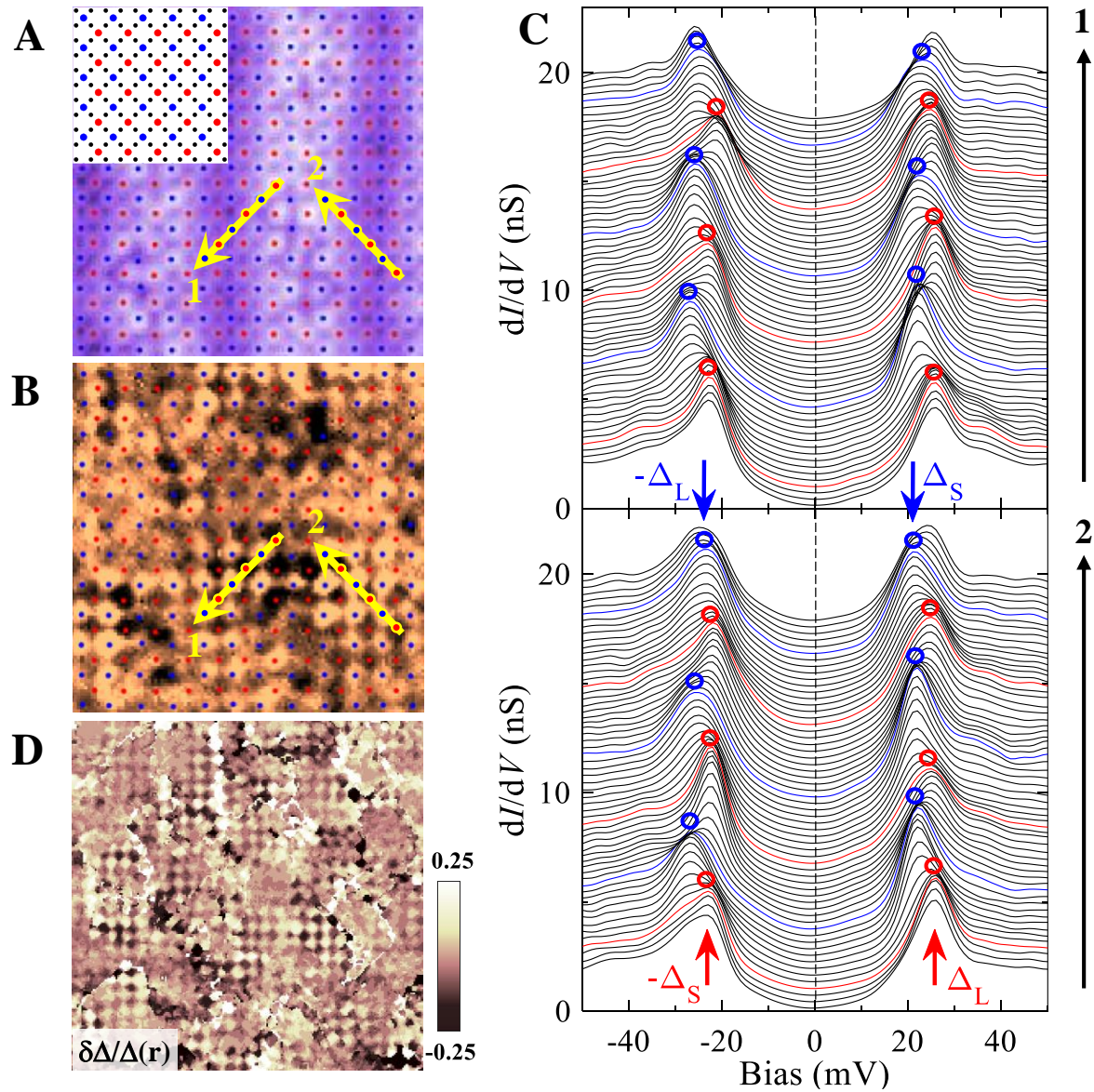

Figure 3

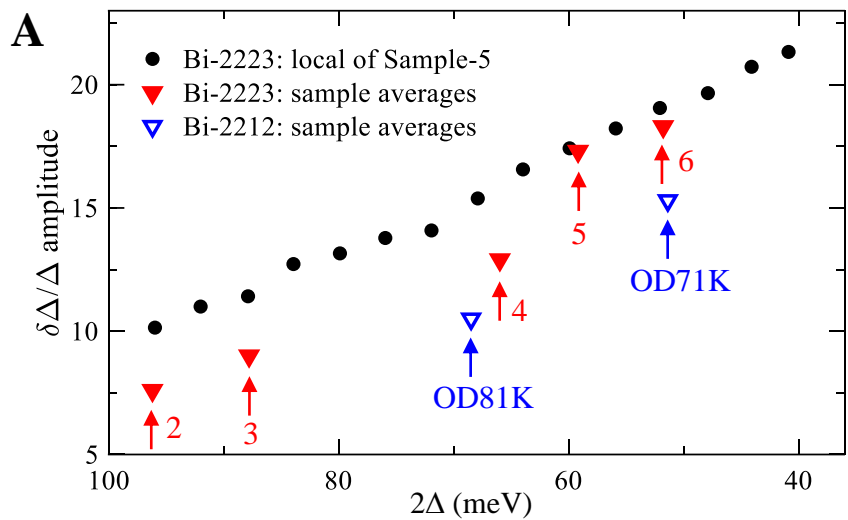
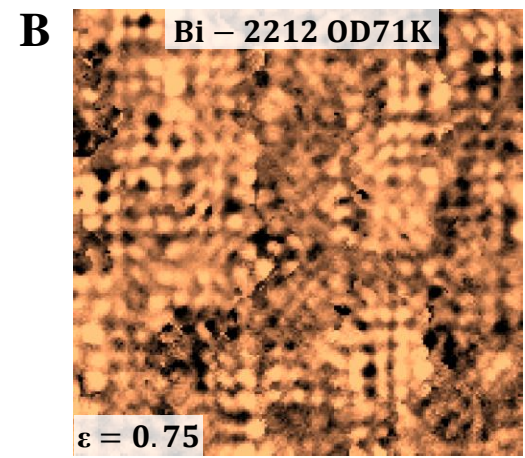
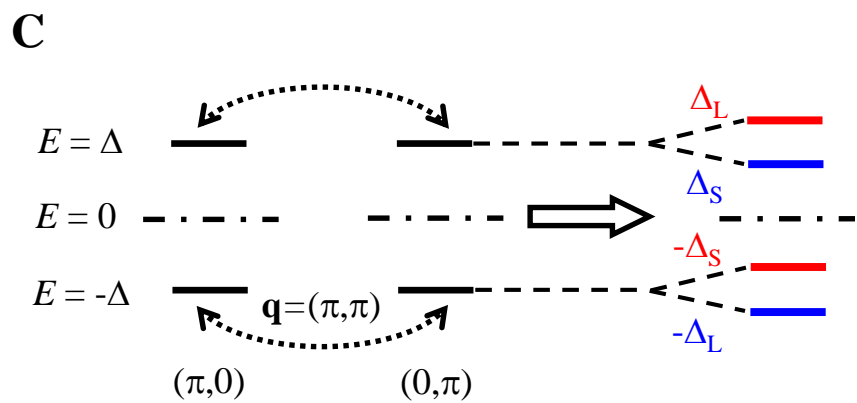
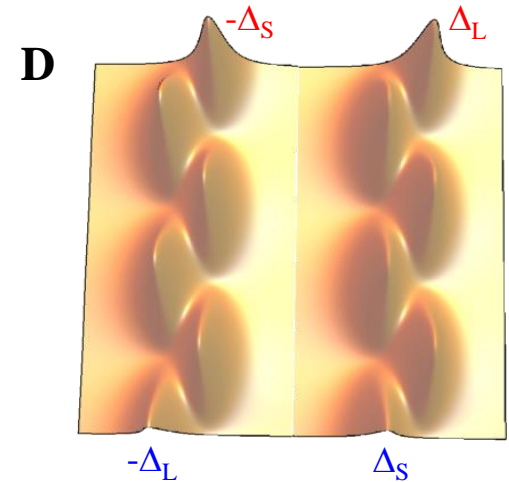

**Figure 4**